\documentclass{ws-rv9x6} 
\usepackage{subfigure}   
\usepackage{ws-rv-thm}   
\usepackage{ws-rv-van}   
\usepackage{bbold}
\usepackage{color}
\makeindex 
 
\begin{document} 
 
\chapter[Effective operators for valence space calculations] {Effective operators for valence space calculations \\ 
from the {\itshape ab initio} No-Core Shell Model} 
 
\author{Zhen Li, N. A. Smirnova}
\address{LP2IB (CNRS/IN2P3 -- Universit\'e de Bordeaux), 33175 Gradignan cedex, France }

\author{A. M. Shirokov} 
\address{Skobeltsyn Institute of Nuclear Physics, Lomonosov Moscow State University, Moscow 119991, Russia} 

\author{I. J. Shin} 
\address{Rare Isotope Science Project, Institute for Basic Science, Daejeon 34000, Republic of Korea} 

\author{B. R. Barrett}  
\address{Department of Physics, University of Arizona, Tucson, Arizona 85721, USA}

\author[Zhen Li, N. A. Smirnova, A. M. Shirokov et al.]{P. Maris, J. P. Vary}  
\address{Department of Physics and Astronomy, Iowa State University, Ames, Iowa 50011, USA} 
 
\begin{abstract} 

In recent years, remarkable progress has been achieved in developing 
novel non-perturbative techniques for constructing 
valence space shell model Hamiltonians from realistic internucleon interactions.
One of these methods is based on the Okubo--Lee--Suzuki (OLS) unitary transformation applied to
no-core shell model (NCSM) solutions.
In the present work, we 
implement the corresponding approach to solve for valence space effective electromagnetic operators.
To this end, we use the NCSM results for $A=16-18$, obtained at $N_{\rm max}=4$, to derive 
a charge-dependent version of the effective interaction for the $sd$ shell, which allows us to exactly reproduce
selected NCSM spectra of $^{18}$O, $^{18}$F and $^{18}$Ne within the two valence nucleon space. 
We then deduce effective single-particle matrix elements of electric quadrupole ($E2$) and 
magnetic dipole ($M1$) operators by matching them to the electromagnetic transitions and 
moments for $^{17}$O and $^{17}$F from the NCSM at $N_{\rm max}=4$.
Thus, effective $E2$ and $M1$ operators are obtained as sets of   
single-particle matrix elements for  the valence space ($sd$ shell) which allow us to reproduce
the NCSM results for $A=17$ exactly. 
Systematic comparison of a large set of $sd$ shell results on quadrupole and 
magnetic dipole moments and transitions for $A=18$ using effective $E2$ and $M1$ 
operators that we derive from the full NCSM calculations demonstrates a remarkable agreement.


\end{abstract} 
\body 
 

\section{Preamble}

{\em The authors wish to express their appreciation for the opportunity to participate in this honorary volume for Prof. Akito Arima.  
Indeed, Prof. Arima led many pioneering works in the microscopic theory of nuclear forces and effective interactions.  
Several of the authors had the great pleasure to discuss these fundamental physics topics with him and to receive his sage advice. 
With this contribution, we propose extensions to microscopic theory of valence interactions including electromagnetic effective 
operators that follow lines of thinking that can trace their origin to the fundamental works of Prof. Akito Arima.}

\section{Introduction} 

Outstanding progress in nuclear many-body computations and in construction of highly accurate realistic internucleon interactions
during the last couple of decades has made it possible to describe light nuclei and intermediate-mass nuclei 
near closed shells from first principles~\cite{Barrett2013,Hagen2014, Carlson2015,Hergert2016,Dickhoff2004,Tichai2020,Maris2021}.
Among all these {\em ab initio}  approaches, 
the no-core shell model~\cite{Barrett2013} (NCSM), being a full configuration-interaction type method,
solves the eigenvalue problem by exact diagonalization of the Hamiltonian matrix constructed in a large-dimensional 
many-body harmonic-oscillator basis. 
The latest state-of-the-art computations, such as those made possible by the Many-fermion dynamics code for nuclei (MFDn)~\cite{MFDn1,MFDn2,MFDn3,MFDn4,MFDn5}, 
are capable of providing converged or nearly-converged  solutions for nuclei up to $A\sim 18$, 
setting benchmarks for other nuclear structure models~\cite{Maris2021,SaSr2020,Nmax=6}.
However, for heavier nuclei the eigenproblem has still to be solved in a more restricted model space, 
typically for a few valence nucleons occupying one or two harmonic oscillator shells beyond a closed-shell core.
This approach with an assumed core of passive nucleons is  known as a traditional interacting shell-model (
see, e.\:g., Refs.~\cite{Brown2001,OtHo2001,CaurierRMP}).
Severe truncation of the model space requires construction of an appropriate {\em effective Hamiltonian} and
{\em effective electroweak operators}.
From the 1960's, for more than fifty years, the many-body perturbation theory (
see, e.\:g., Refs.~\cite{MHJ95,Stroberg2019} and references therein) was 
the dominant approach to deal with the microscopic derivation of effective Hamiltonians.
It was, however, noticed that when derived from two-nucleon interaction, an effective valence-space Hamiltonian could not compete in precision with empirically obtained effective Hamiltonians~\cite{USD,GXPF1A}, mainly because of its deficient monopole part~\cite{PoZu81}. One of possible reasons was identified as the absence of three-nucleon forces~\cite{Zuker3N}.

New non-perturbative techniques for deriving effective valence-space interactions have been developed during the last decade, such as
similarity-renormalization group (SRG) inspired methods~\cite{StrobergPRL2017}, or application of Okubo--Lee--Suzuki (OLS)
unitary transformation~\cite{Okubo1954,Suzuki1980,Suzuki1982} to either a NCSM solution~\cite{Lisetskiy2009,Dikmen2015,SmiBa2019} or 
to a solution found within the coupled cluster theory~\cite{Jansen2016,Sun2018}.  
Recent inclusion of three-nucleon forces via different approaches \cite{Otsuka3N,Holt2012,Jansen2016,StrobergPRL2017,Fukui2018,Stroberg2019}
seems to cure the deficiency of the monopole part of the effective interactions and 
brings encouraging improvement to the theoretical description.

Construction of appropriate effective electroweak operators to be used in a valence space has been another long-standing challenge.
Since the earlier work by Arima and collaborators~\cite{Arima1954a,Arima1954b}, the importance of accounting for the 
restricted model space has been realized. 
Many efforts were devoted to the construction of effective electromagnetic and Gamow--Teller operators  both 
microscopically~\cite{Kuo1973,Brown1977,Oset1979,Arima1987,Towner1987,Parzuchowski2017} 
or empirically by a fit of the 
optimum form of a corresponding operator to the data~\cite{Brown1982,Carchidi1986,USD,USDab,USDab_em}.

In this work, we use the NCSM solutions for $A=16{-}18$ nuclei
 in order to construct an effective
interaction and effective $E2$ and $M1$ operators for the valence $sd$-shell model space.
We construct here a charge-dependent version of the effective interaction presented earlier~\cite{SmiBa2019}
based on the Daejeon16 nucleon-nucleon ($NN$) potential~\cite{DJ16}, using
$^{18}$O, $^{18}$F and $^{18}$Ne to derive neutron-neutron, neutron-proton and proton-proton
two-body matrix elements (TBMEs). 
Effective $E2$ and $M1$ single-particle matrix elements are obtained by matching them to the corresponding 
NCSM results for $^{17}$O and $^{17}$F 
obtained with basis spaces including all many-body excitation oscillator quanta up to $N_{\rm max}=4$. 
These constructed effective operators are 
tested on a large number of transitions and moments
for $A=18$ nuclei.

\section{Microscopic effective interaction from the NCSM} 

The derivation of valence-space effective interactions from the NCSM is described in detail in Refs.~\cite{Dikmen2015,SmiBa2019}{}
and references therein.
We start with a translationally-invariant non-relativistic Hamiltonian for $A$ point-like nucleons,
containing relative kinetic energies and $NN$ interactions:
\begin{equation}
\label{H_intrinsic}
H =\sum_{i<j}^{A}\frac{(\vec{p_i}- \vec{p_j})^2}{2mA}+\sum_{i<j}^{A}V_{ij}^{NN}.
\end{equation}
Here $m$ is the nucleon mass  (approximated here as the average of the neutron and proton mass), 
$\vec{p}_i$ are nucleonic momenta and $V_{ij}^{NN}$ denotes the bare $NN$ interaction. 
To our chosen Daejeon16 $NN$ interactions we have added the two-body Coulomb interaction between the protons.

The eigenproblem for $H$ is solved by diagonalizing the Hamiltonian matrix, 
constructed in a many-body spherical harmonic-oscillator basis, using the MFDn code~\cite{MFDn1,MFDn2,MFDn3,MFDn4,MFDn5}. 
All nucleons are considered as active particles and are treated equally.
The model space is defined by two parameters: (i) by a given  oscillator 
quantum, $\hbar \Omega $, and 
(ii) by the maximum total many-body 
harmonic-oscillator excitation quanta, $N_{\rm max}$. 
This means that we retain only many-body configurations satisfying the condition 
$\sum_{i=1}^A (2n_i+l_i) \le N_{\rm min} + N_{\rm max}$,
where $n_i$ is the single-particle radial harmonic-oscillator quantum number, 
$l_i$ is the single-particle orbital angular momentum quantum number, while 
$N_{\rm min}$ is the minimal many-body oscillator quanta satisfying 
the Pauli principle for the chosen nucleus.
The use of the harmonic-oscillator basis allows us to remove states with center-of-mass 
excitations by using the Lawson method~\cite{Lawson}.

In the present study, we select $\hbar \Omega =14$~MeV, which is close to the empirical value~\cite{Kirson07} in the vicinity of $A=16$,
and which was also  adopted in our previous work~\cite{SmiBa2019}.


In principle, one may apply the OLS transformation \cite{Okubo1954,Suzuki1980,Suzuki1982} to a ``cluster'' of nucleons 
to develop an effective interaction for the chosen NCSM space (the ``$P$-space'')\cite{Barrett2013}. 
This is referred to as the cluster approximation.  A compact presentation of the OLS method is presented in Ref.~\cite{Vary2018}. 
The goal is to achieve an effective interaction at fixed cluster size (conventionally two-nucleon or three-nucleon) 
for which the NCSM calculations converge more rapidly with increasing $P$-space 
than using the original (non-transformed) interaction. 

In this work we adopt the Daejeon16 $NN$ interaction~\cite{DJ16} which is based on a chiral N3LO interaction~\cite{EnMa2002,EnMa2003}
and which has been softened already by an SRG transformation~\cite{SRG1,SRG2} that improves convergence. 
Hence, following our procedure~\cite{SmiBa2019} we elect to use Daejeon16 directly, 
without an additional OLS transformation for our NCSM calculations.

For this work, we adopt the NCSM calculations at $N_{\rm max}$ = 4 for the $A=16$, 17 and $^{18}$F.  
In addition, to obtain a charge-dependent valence effective interaction for the first time, 
we perform new NCSM calculations at $N_{\rm max}$ = 4 for $^{18}$O and $^{18}$Ne.

The resulting NCSM eigenvalues and eigenvectors are then employed to define a new effective interaction 
for the $N_{\rm max} = 0$ space (the ``$P'$-space'') through the   application of the OLS method.  
The NCSM results dominated by $sd$-shell configurations are appropriate for adoption to define the $P'$-space. 
That is, from the NCSM results for $A = 16{-}18$ we apply the OLS method to obtain an effective valence space 
($sd$-shell) Hamiltonian consisting of a core, a single-particle and a two-particle interaction~\cite{Dikmen2015,SmiBa2019}.
This effective interaction is guaranteed to reproduce the NCSM results for the $^{16}$O core and the states of $A = 17$ and 18 
selected to be included in the definition of the $P'$-space.  


From the $A = 18$ NCSM results  we select the number of eigenpairs (eigenvalues and eigenvectors) for each $J$ governed by the number of needed two-body matrix elements. 
Let us label the total number of eigenpairs (adding up the dimensions for each $J$) as $d_{nn}$, $d_{pp}$ and $d_{np}$ for valence space configurations of 
$^{18}$O, $^{18}$Ne and $^{18}$F respectively.
These are the states, selected from among the NCSM candidates, which have the largest valence space ($0\hbar \Omega $ in the present study) probability.
It should be noted that these are not necessarily the states with the lowest NCSM eigenvalues.
The exact dimensions are $d_{nn}=14$ for  $^{18}$O, $d_{pp}=14$ for $^{18}$Ne,
and $d_{np}=28$ for $^{18}$F, since it includes both $T=0$ and $T=1$ states.

\begin{table}[htb]
\tbl{Neutron-neutron, proton-neutron ($T=1$) and proton-proton TBMEs of the derived effective interaction $\langle ab;J,T=1 | V_{\rm eff} | cd; J,T=1 \rangle $ in the $sd$ shell
from Daejeon16. Note that the Coulomb interaction is included in the NCSM calculations. See text for details.}
{\begin{tabular}{rrrrrrrrr}
\toprule
$2j_a$ & $2j_b$ & $2j_c$ & $2j_d$ & $J$ & $T$ & $V_{nn}$ & $V_{pn}^{T=1}$  & $V_{pp}$\\
\hline
   1 &   1 &   1 &   1 &   0 &   1 & $   -2.018$ & $   -1.989$ &  $   -1.536$ \\
   1 &   1 &   3 &   3 &   0 &   1 & $   -0.713$ & $   -0.769$ &  $   -0.797$ \\
   1 &   1 &   5 &   5 &   0 &   1 & $   -1.446$ & $   -1.422$ &  $   -1.344$ \\
   3 &   3 &   3 &   3 &   0 &   1 & $   -1.350$ & $   -1.317$ &  $   -0.847$ \\
   3 &   3 &   5 &   5 &   0 &   1 & $   -2.377$ & $   -2.500$ &  $   -2.582$ \\
   5 &   5 &   5 &   5 &   0 &   1 & $   -2.618$ & $   -2.590$ &  $   -2.046$ \\
   1 &   3 &   1 &   3 &   1 &   1 & $    0.137$ & $    0.136$ &  $    0.482$ \\
   1 &   3 &   3 &   5 &   1 &   1 & $   -0.151$ & $   -0.151$ &  $   -0.152$ \\
   3 &   5 &   3 &   5 &   1 &   1 & $    0.069$ & $    0.062$ &  $    0.441$ \\
   1 &   3 &   1 &   3 &   2 &   1 & $   -0.500$ & $   -0.500$ &  $   -0.116$ \\
   1 &   3 &   1 &   5 &   2 &   1 & $   -1.519$ & $   -1.498$ &  $   -1.438$ \\
   1 &   3 &   3 &   3 &   2 &   1 & $   -0.090$ & $   -0.099$ &  $   -0.089$ \\
   1 &   3 &   3 &   5 &   2 &   1 & $    0.459$ & $    0.452$ &  $    0.428$ \\
   1 &   3 &   5 &   5 &   2 &   1 & $   -1.183$ & $   -1.164$ &  $   -1.120$ \\
   1 &   5 &   1 &   5 &   2 &   1 & $   -1.358$ & $   -1.341$ &  $   -0.908$ \\
   1 &   5 &   3 &   3 &   2 &   1 & $   -0.854$ & $   -0.853$ &  $   -0.828$ \\
   1 &   5 &   3 &   5 &   2 &   1 & $    0.366$ & $    0.360$ &  $    0.336$ \\
   1 &   5 &   5 &   5 &   2 &   1 & $   -0.549$ & $   -0.542$ &  $   -0.509$ \\
   3 &   3 &   3 &   3 &   2 &   1 & $   -0.187$ & $   -0.185$ &  $    0.180$ \\
   3 &   3 &   3 &   5 &   2 &   1 & $    0.861$ & $    0.855$ &  $    0.824$ \\
   3 &   3 &   5 &   5 &   2 &   1 & $   -0.818$ & $   -0.837$ &  $   -0.836$ \\
   3 &   5 &   3 &   5 &   2 &   1 & $   -0.108$ & $   -0.116$ &  $    0.267$ \\
   3 &   5 &   5 &   5 &   2 &   1 & $    0.252$ & $    0.259$ &  $    0.263$ \\
   5 &   5 &   5 &   5 &   2 &   1 & $   -1.140$ & $   -1.132$ &  $   -0.695$ \\
   1 &   5 &   1 &   5 &   3 &   1 & $    0.435$ & $    0.433$ &  $    0.799$ \\
   1 &   5 &   3 &   5 &   3 &   1 & $    0.310$ & $    0.302$ &  $    0.289$ \\
   3 &   5 &   3 &   5 &   3 &   1 & $    0.186$ & $    0.169$ &  $    0.508$ \\
   3 &   5 &   3 &   5 &   4 &   1 & $   -1.628$ & $   -1.621$ &  $   -1.196$ \\
   3 &   5 &   5 &   5 &   4 &   1 & $    1.388$ & $    1.371$ &  $    1.323$ \\
   5 &   5 &   5 &   5 &   4 &   1 & $   -0.212$ & $   -0.211$ &  $    0.184$ \\
\botrule 
\end{tabular} 
}
\label{tab:TBMEs} 
\end{table} 

\begin{table}[htb]
\tbl{Proton-neutron ($T=0$) TBMEs of the derived effective interaction $\langle ab;J,T=0 | V_{\rm eff} | cd; J, T=0 \rangle $ in the $sd$ shell
from Daejeon16. See text for details.}
{\begin{tabular}{rrrrrrrrr}
\toprule
$2j_a$ & $2j_b$ & $2j_c$ & $2j_d$ & $J$ & $T$ & $V_{pn}^{T=0}$ \\
\hline
   1 &   1 &   1 &   1 &   1 &   0 &  $   -2.909$ \\
   1 &   1 &   1 &   3 &   1 &   0 &  $    0.381$ \\
   1 &   1 &   3 &   3 &   1 &   0 &  $    0.573$ \\
   1 &   1 &   3 &   5 &   1 &   0 &  $   -2.196$ \\
   1 &   1 &   5 &   5 &   1 &   0 &  $   -1.224$ \\
   1 &   3 &   1 &   3 &   1 &   0 &  $   -3.565$ \\
   1 &   3 &   3 &   3 &   1 &   0 &  $    2.176$ \\
   1 &   3 &   3 &   5 &   1 &   0 &  $    1.458$ \\
   1 &   3 &   5 &   5 &   1 &   0 &  $   -0.761$ \\
   3 &   3 &   3 &   3 &   1 &   0 &  $   -1.343$ \\
   3 &   3 &   3 &   5 &   1 &   0 &  $    0.305$ \\
   3 &   3 &   5 &   5 &   1 &   0 &  $    2.465$ \\
   3 &   5 &   3 &   5 &   1 &   0 &  $   -5.702$ \\
   3 &   5 &   5 &   5 &   1 &   0 &  $   -3.375$ \\
   5 &   5 &   5 &   5 &   1 &   0 &  $   -1.043$ \\
   1 &   3 &   1 &   3 &   2 &   0 &  $   -1.839$ \\
   1 &   3 &   1 &   5 &   2 &   0 &  $    3.045$ \\
   1 &   3 &   3 &   5 &   2 &   0 &  $   -2.290$ \\
   1 &   5 &   1 &   5 &   2 &   0 &  $   -0.497$ \\
   1 &   5 &   3 &   5 &   2 &   0 &  $    1.887$ \\
   3 &   5 &   3 &   5 &   2 &   0 &  $   -3.370$ \\
   1 &   5 &   1 &   5 &   3 &   0 &  $   -3.935$ \\
   1 &   5 &   3 &   3 &   3 &   0 &  $   -0.004$ \\
   1 &   5 &   3 &   5 &   3 &   0 &  $   -1.537$ \\
   1 &   5 &   5 &   5 &   3 &   0 &  $   -2.011$ \\
   3 &   3 &   3 &   3 &   3 &   0 &  $   -3.144$ \\
   3 &   3 &   3 &   5 &   3 &   0 &  $   -1.743$ \\
   3 &   3 &   5 &   5 &   3 &   0 &  $    0.977$ \\
   3 &   5 &   3 &   5 &   3 &   0 &  $   -0.646$ \\
   3 &   5 &   5 &   5 &   3 &   0 &  $   -2.071$ \\
   5 &   5 &   5 &   5 &   3 &   0 &  $   -0.690$ \\
   3 &   5 &   3 &   5 &   4 &   0 &  $   -4.218$ \\
   5 &   5 &   5 &   5 &   5 &   0 &  $   -4.448$ \\
\botrule 
\end{tabular} 
}
\label{tab:TBMEs_T=0} 
\end{table} 
 
The OLS transformation leads to a two-nucleon effective Hamiltonian defined in the $N_{\rm max}=0$ model space by its
TBMEs which in the angular momentum $J$ and isospin $T$ coupled form read $\langle ab;JT | H_{\rm eff} | cd; JT \rangle $,
where $a$, $ b$, $c$, $d$ denote a full set of single-particle spherical quantum numbers which label an orbital, e.g. $a=(n_a,l_a,j_a,t_{z_a})$,
with $t_z$ distinguishing protons and neutrons.

To obtain the one-body neutron and proton contributions (single-particle energies, $\epsilon $), we subtract the ground-state energy of $^{16}$O 
(core energy, $E_{\rm core}$) from $^{17}$O and $^{17}$F, respectively.
To get the TBMEs of the effective interaction in the valence space, we have to subtract
this core energy  and the one-body contributions from diagonal TBMEs of the OLS-derived 
effective valence space Hamiltonian $H_{\rm eff}$,
\begin{equation}
\langle ab;JT | V_{\rm eff} | ab; JT \rangle = \langle ab;JT | H_{\rm eff} | ab; JT \rangle -E_{\rm core} - (\epsilon_a+ \epsilon_b),
\end{equation}
while for non-diagonal TBMEs, we have
\begin{equation}
\langle ab;JT | V_{\rm eff} | cd; JT \rangle = \langle ab;JT | H_{\rm eff} | cd; JT \rangle \,.
\end{equation}
In this way, from NCSM calculations for $^{18}$O, $^{18}$Ne and $^{18}$F, we derive a set of neutron-neutron, proton-proton and proton-neutron TBMEs,
respectively, to be used in the valence space shell-model calculations. 

%
%
%

\begin{figure}[t!]  
\centering{
  \includegraphics[width=4in]{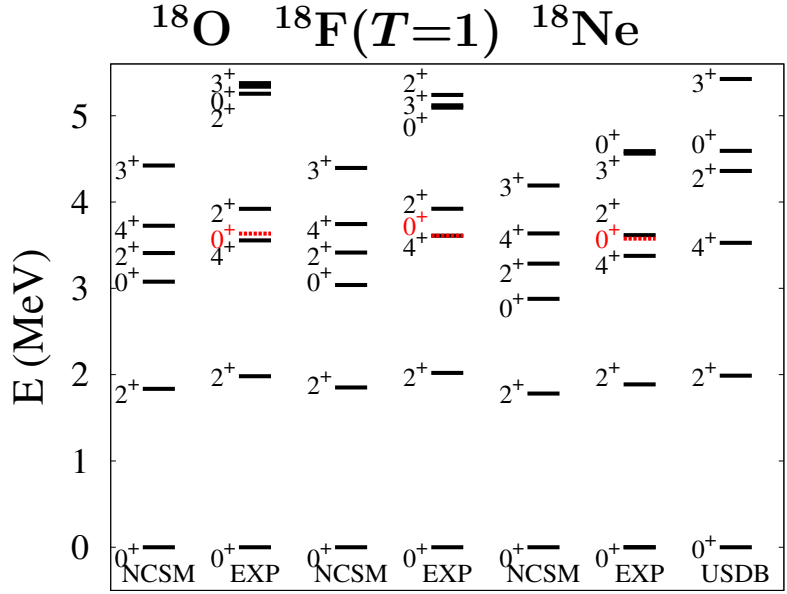}
}
\caption{Low-energy theoretical excitation spectra of $^{18}$O, $^{18}$Ne and the $T=1$ part 
of the $^{18}$F spectrum
obtained within the NCSM with Daejeon16 at $N_{\rm max}=4$ 
(identical to the $sd$-shell calculation with the proposed effective valence-space interaction).  
The USDB $T=1$ spectrum for $A=18$ and low-energy experimental spectra of positive-parity states~\protect\cite{nndc} are shown for comparison. 
The $0^+_2$ state shown in red in experimental spectrum in $^{18}$O
is most probably an intruder state as evidenced by the electromagnetic transition pattern~\cite{nndc}. 
Its isobaric analogue states in $^{18}$F and  $^{18}$Ne are also marked in red, and, therefore, should not be compared
with nearby theoretical spectra.\label{fig:O18_F18_Ne18} } 
\end{figure}

The resulting core energy, proton and neutron single-particles energies, as well as $T=0,1$ proton-neutron TBMEs obtained from Daejeon16
have been published in an earlier study (see column ``bare'' of Table II and column ``bare'' of Table XI in Ref. ~\cite{SmiBa2019}), 
while the neutron-neutron
and proton-proton TBMEs obtained specifically for this work. For the convenience of the reader, we present the whole set of TBMEs in Tables~\ref{tab:TBMEs} and \ref{tab:TBMEs_T=0}.
By construction, the $sd$-shell spectra of $^{18}$O, $^{18}$Ne and  $^{18}$F obtained with the derived effective Hamiltonian
exactly reproduce the results of the NCSM for the chosen $P'$-space eigenvalues. Comparison of the low-energy $T=1$ states
are shown in Fig.~\ref{fig:O18_F18_Ne18} in comparison with experiment~\cite{nndc} and with the results from a phenomenological interaction USDB~\cite{USDab}.
The latter produces identical $T=1$ spectra for all three $A=18$ isobars, because the USDB Hamiltonian is charge-independent, 
including the single-particle energies which are the same for protons and neutrons.

Since there are only small differences among the corresponding proton-proton, neutron-neutron and proton-neutron ($T=1$) TBMEs, as seen from Table~\ref{tab:TBMEs},
(aside from the expected, approximately constant, diagonal Coulomb shift for the proton-proton TBMEs),
the excitation spectra support the approximate validity of isospin symmetry (see Fig.~\ref{fig:O18_F18_Ne18}).

\begin{figure}[t] 
\centerline{   
  \includegraphics[width=4in]{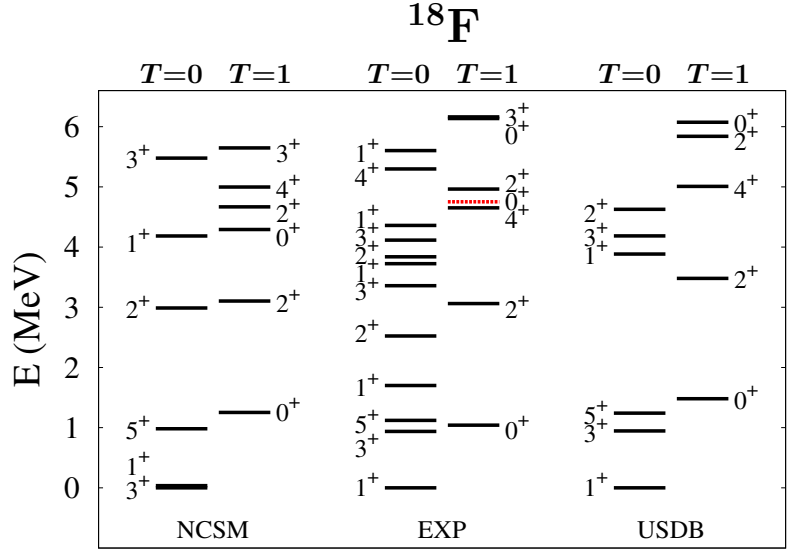}
} 
\caption{Low-energy excitation spectra of $^{18}$F (up to 6.2 MeV): theoretical spectrum
obtained within the NCSM with Daejeon16 at $N_{\rm max}=4$ 
(identical to the $sd$-shell calculation with the effective interaction of Ref.~\cite{SmiBa2019}), 
in comparison with the positive-parity experimental spectrum\protect\cite{nndc}  and the one obtained from USDB.
The $0^+_2$, $T=1$ state in $^{18}$F, analogue to the intruder state in $^{18}$O, is shown in red color.\label{fig:F18} }
\end{figure}

The low-lying $T=0$ spectrum of $^{18}$F,
obtained from the NCSM, contains states with the same $J^{\pi }$ quantum numbers as the USDB spectrum, however, 
it is more extended in energy than the USDB spectrum, with different ordering,
and the lowest $3^+$ state, being almost degenerate with $1^+$ state, becomes the ground state (see Fig.~\ref{fig:F18}). 
Except for the three lowest states, the experimental $T=0$ spectrum is noticeably more dense and it is challenging
to find theoretical counterparts in the $sd$ shell calculation. 
This may hint at a possible intruder character of some of those states.
In the NCSM calculation at $N_{\rm max}=4$, states dominated by $2\hbar \Omega $ configurations are still located at much 
higher energy and therefore too far from being converged to shed light on this intruder state issue. 
Since the present study is focused on the construction of effective operators and to the comparison between the NCSM
and valence space results, we do not investigate potential intruder states in the current work.
In future applications with larger $P$-spaces, one may anticipate, however, that as these intruders fall further into the region of the valence-space dominated states, the mixing that is likely to occur will lead to complications in identifying the appropriate eigenpairs for retention in the $P'$ space.

\section{Effective electromagnetic operators  in the $sd$ shell}

Considering nucleons as non-relativistic point-like particles and assuming a long-wavelength approximation for emitted or absorbed radiation, 
one can deduce the following expressions for one-body electric quadrupole and magnetic dipole operators:
\begin{equation}
\hat O(E2) = \sum_{k=1}^A e_k r_k^2 \hat Y_2(\hat r_k) \label{E2}\,,
\end{equation}
and
\begin{equation}
\hat O(M1) = \sum_{k=1}^A \mu_N \left[ g^s_k \vec{s}_k +  g^l_k \vec{l}_k\right]\label{M1}\,,
\end{equation}
where $\hat Y_2(\hat r_k)$ are spherical harmonics of rank $2$, $\hat r_k = (\theta_k, \phi_k)$ being a spherical angle,
$\vec{s}_k$ and $\vec{l}_k$ are spin and orbital  angular momenta operators of a nucleon, respectively,
$e_k$ denotes the electric charge of the $k$th nucleon, with the bare values being $e_p=e$ for a proton and $e_n=0$ for a neutron,
while $g^s$ and $g^l$ are spin and orbital gyro-magnetic factors ($g$-factors) with bare values being
$g^s_p =5.586$, $g^l_p =1$, $g^s_n =-3.826$, $g^l_n =0$ for protons and neutrons, respectively.
For valence space calculations, $e_p$ and $e_n$ are conventionally understood to be {\em effective} proton and neutron electric charges,
while $g^{l,s}_k$ have to be {\em effective} $g$-factors.
In principle one could follow the same procedure as we use to develop the valence space effective two-nucleon interaction to develop valence space effective $E2$ and $M1$ operators.  
The procedures for doing so would follow those demonstrated in Refs.~\cite{Vary2018,Stetcu2005}. 
The results, for the $P'$-space we define here, would be represented as TBMEs for these operators.  
That is, there are, in theory, both one-body and two-body contributions to such derived effective valence electromagnetic operators. 
Here, we will obtain approximate theoretical one-body effective valence space operators for $E2$ and $M1$.  
We then test those approximate operators in a number of ways.

Our starting point is the NCSM calculation of $E2$ and $M1$ transitions and moments in $^{17}$O and $^{17}$F
at $N_{\rm max}=4$. The three lowest states are $1/2^+$, $5/2^+$ and $3/2^+$ states which are mapped into single-particle
$1s_{1/2}$, $0d_{5/2}$ and $0d_{3/2}$ states in the $sd$-shell calculation. 
Therefore, we require that the matrix elements of the $E2$ and $M1$ operator with bare electric charges and $g$-factors 
between the many-body eigenstates from the NCSM at $N_{\rm max}=4$ 
be exactly reproduced by effective single-particle matrix elements in the valence space.

\subsection{Effective $E2$ operator}

To obtain an effective one-body valence-space $E2$ operator, 
we use the NCSM results for $E2$ transitions and quadrupole moments for $^{17}$O and $^{17}$F.
For $^{17}$F, we define a valence-state-dependent  proton effective charge such that it exactly reproduces
the NCSM values for transitions and moments.
Next, we develop a state-dependent effective charge for the valence neutron.  
We note that the NCSM uses bare electric charges, so only protons contribute to the matrix elements of the $E2$ operator.
In contrast, in the valence space $^{17}$O has only one valence neutron and therefore we require that it carries an effective
charge to be able to reproduce electromagnetic transitions and moments by the valence space calculations.
These rules can be summarized as follows:
\begin{equation}
\begin{array}{ll}
e_n(a,b) \langle b || r^2 \hat Y_2(\hat r )  || a\rangle =\langle J_f || \hat O(E2) || J_i \rangle & ({\rm from} \; ^{17}{\rm O}) \\ 
e_p(a,b) \langle b || r^2 \hat Y_2(\hat r )  || a\rangle =\langle J_f || \hat O(E2) || J_i \rangle & ({\rm from} \; ^{17}{\rm F}) \\ 
\end{array}
\label{neffch}
\end{equation} 
where $e_n(a,b)$ and $e_p(a,b)$ are neutron and proton effective state-dependent charges, $J_i$ and $J_f$ are the initial and final state obtained within the NCSM,
which are mapped onto single-particle states $a$ and $b$, while the $E2$ operator $\hat O(E2)$ is defined by Eq.~(\ref{E2}).
Double bars in this equation and below refer to the matrix element reduced  in angular momentum.

\begin{figure}[!] 
     \centerline{\includegraphics[width=3.7in]{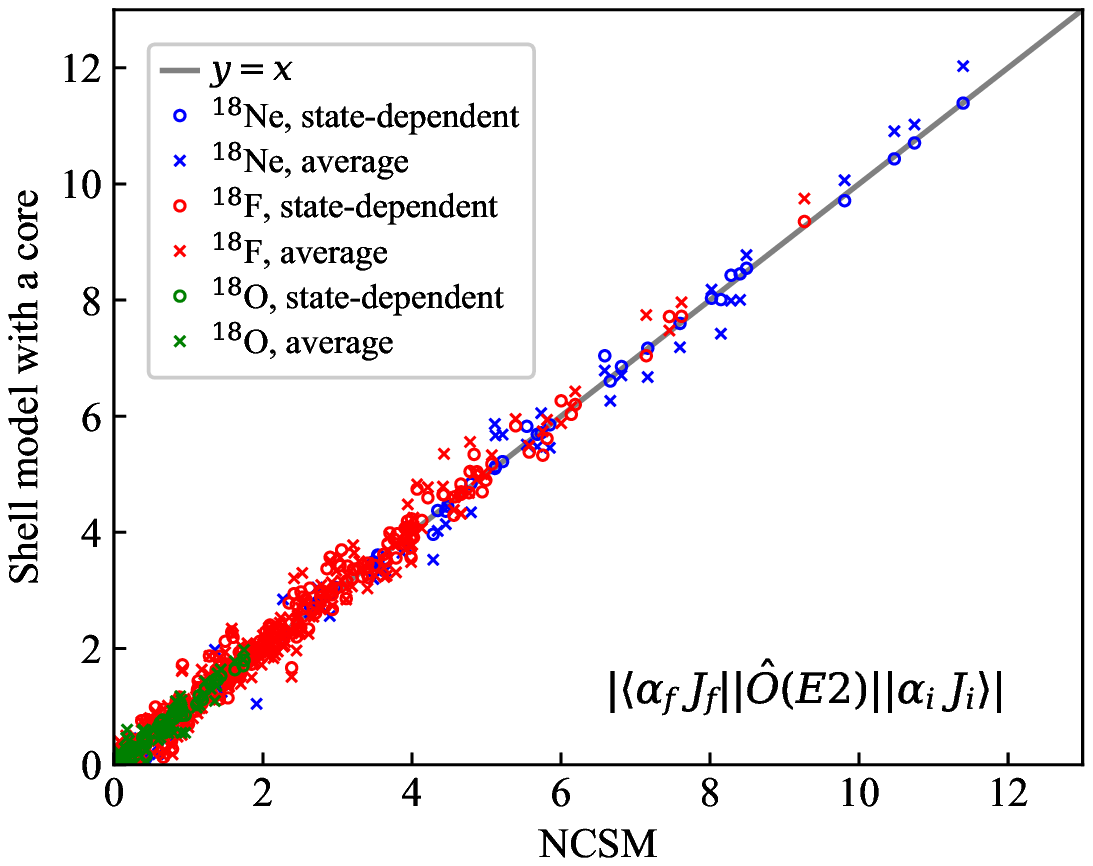}} 
     \centerline{\includegraphics[width=3.7in]{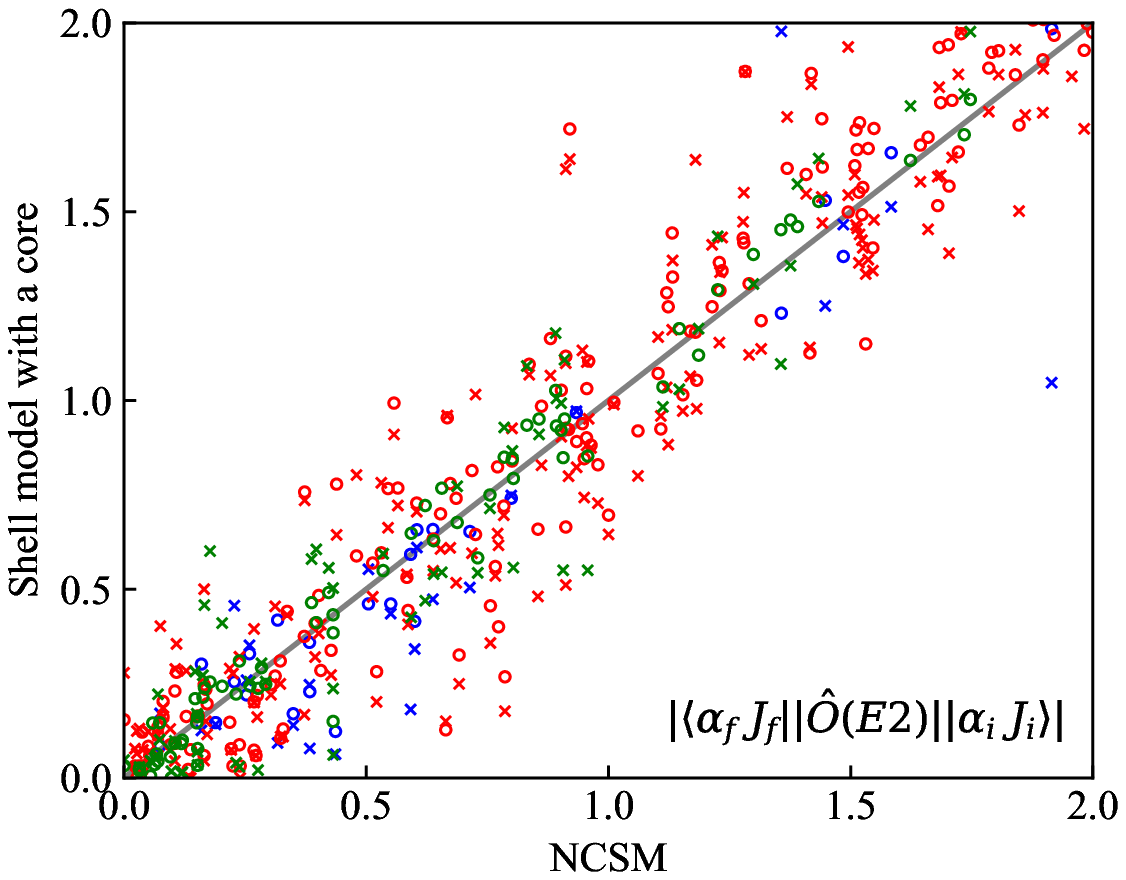}}
\caption{Comparison of the $sd$-shell results for the absolute values of reduced $E2$ transition matrix elements 
between various states in $^{18}$O, $^{18}$F and $^{18}$Ne obtained using effective $E2$ operators  within shell model 
with a core with those obtained by the NCSM (top); the same in the vicinity of zero 
in order to provide details on an enlarged scale for the weaker transitions (bottom).}\label{fig:BE2} 
\vspace{-2ex}\end{figure}

\begin{figure}[t] 
\centerline{ 
     {\includegraphics[width=3.7in]{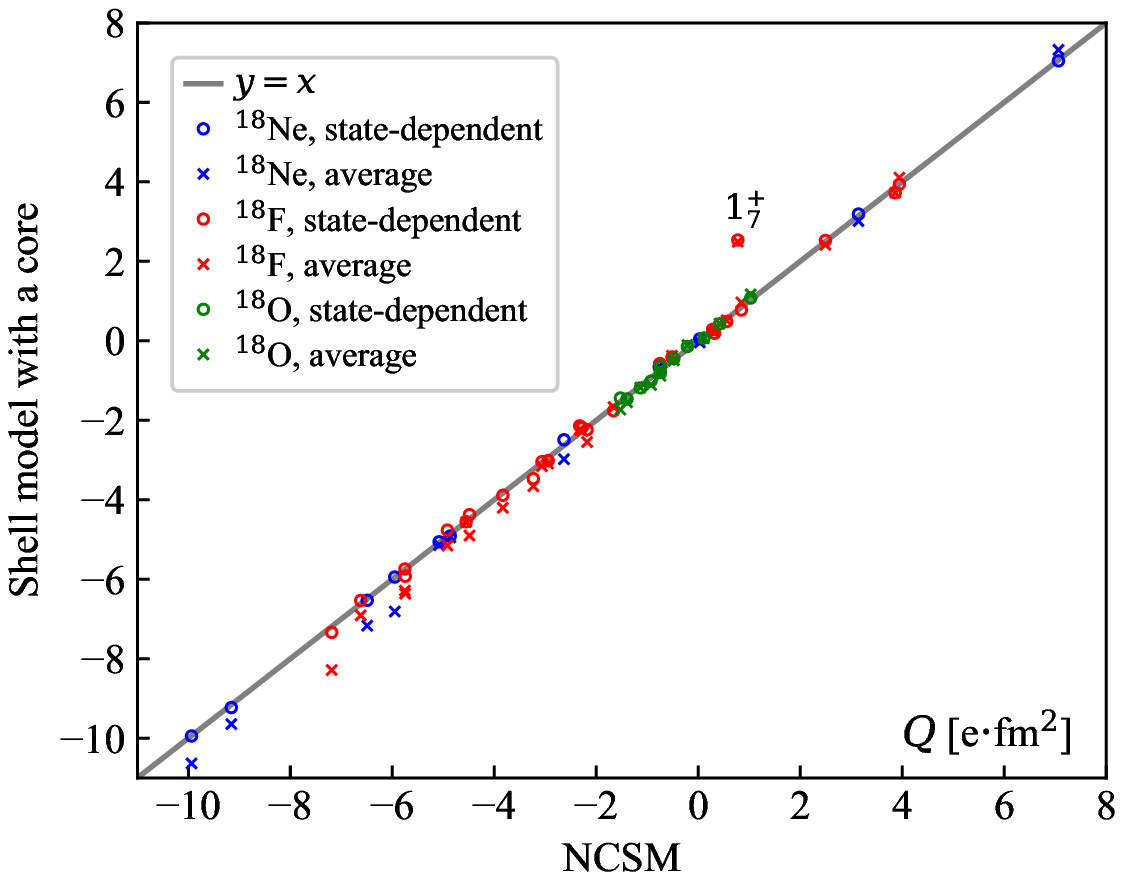}}
} 
\caption{Comparison of the $sd$-shell results for 
electric quadrupole moments of various states in $^{18}$O,
$^{18}$F and $^{18}$Ne obtained using effective $E2$ with those obtained by the NCSM.
An outlier in $^{18}$F, which is discussed in the text, is flagged with its spin and 
sequence in the spectrum indicated.}\label{fig:Q} 
\end{figure}

Matching the NCSM results for $E2$ transitions and quadrupole moments for $^{17}$O and $^{17}$F, 
we have deduced the values for effective electric charges which are summarized in Table~\ref{tab:eff_charges_gfactors}.
These effective charges  $e_{p}(a,b)$ and $e_{n}(a,b)$
depend on the quantum numbers of the initial and final states, $a$ and $b$,
so the effective $E2$ operator is deduced as a set of individually parameterized single-particle matrix elements $\langle b |\hat O(E2)| a \rangle $.
However, the state dependence of the effective charges is not strong as is seen from 
Table~\ref{tab:eff_charges_gfactors}: the difference between neutron effective charges $e_{n}(a,b)$ does
not exceed $\sim 0.12e$ and  the maximal difference between various proton effective charges $e_{p}(a,b)$ 
is $\sim 0.25e$. The weak state dependence of effective charges tends to support the use of  state-independent
effective neutron and proton charges as is conventionally done in the valence-space calculations,
see, e.\:g., Ref.~\cite{USDab_em}{}. We present in Table~\ref{tab:eff_charges_gfactors} 
average theoretical effective neutron, $\bar{e}_{n}$, and proton, $\bar{e}_{p}$,   charges which are seen to be
smaller than the commonly accepted values~\cite{USDab_em} for $0\hbar \Omega $ model space calculations,
$e_p=1.36(5)e$ and $e_n = 0.45(5)e$.

\begin{table}[ht] 
\tbl{Effective charges (in units of $e$) and effective $g$-factors in the $sd$-shell valence space found by  
the respective NCSM calculations for $^{17}$O and $^{17}$F. 
The bare values used in the NCSM calculations are also indicated for reference. 
Hermiticity requires symmetry of these effective quantities under the interchange of $a$ and $b$.
See text for details. }
{\begin{tabular}{@{}lcccccc@{}} \toprule
$( a,b)$  & $e_n(a,b)$ & $e_p(a,b)$    & $g^s_n(a,b)$ & $g^l_n(a,b)$  & $g^s_p(a,b)$ & $g^l_p(a,b)$  \\ \colrule
bare                  & 0.0    & 1.0   & -3.826       & 0.0         & 5.586        & 1.0         \\ \colrule
$(0d_{5/2},1s_{1/2})$ & 0.181  & 1.171 &              &             &              & \\ \colrule
$(0d_{5/2},0d_{3/2})$ & 0.281  & 1.236 & -3.608 & 0.020 & 5.252 & 0.916 \\ \colrule
$(1s_{1/2},0d_{3/2})$ & 0.168  & 1.297 &        &       &       &       \\ \colrule
$(0d_{5/2},0d_{5/2})$ & 0.179  & 1.060 & -3.751 & 0.026 & 5.499 & 0.976 \\ \colrule
$(0d_{3/2},0d_{3/2})$ & 0.172  & 1.248 & -3.690 & 0.033 & 5.332 & 0.957  \\ \colrule
$(1s_{1/2},1s_{1/2})$ &        &       & -3.729 &       & 5.468 &  \\ \colrule
                      & $\overline{e}_n$ & $\overline{e}_p$ & $\overline{g}^s_n$ & $\overline{g}^l_n$ & $\overline{g}^s_p$ & $\overline{g}^l_p$  \\ \colrule
average               & 0.196  & 1.202 & -3.695  & 0.026 & 5.388 & 0.950 \\ \botrule 
\end{tabular}  
}
\label{tab:eff_charges_gfactors} 
\end{table}

To assess the quality of the operator constructed above, we calculate reduced transition matrix elements and quadrupole moments
in $A=18$ $sd$-shell nuclei using  both state-dependent and average effective charges. 
Reduced transition matrix elements, $\langle \alpha_f J_f  || \hat O(E2) || \alpha_i J_i \rangle  $,  are required to express
the corresponding $B(E2)$ values of a transition from
an initial state $| \alpha_i J_i \rangle $ to the final state $| \alpha_f J_f \rangle $ as
\begin{equation}
B(E2;J_i \to J_f) =  \dfrac{1}{2J_i+1} |\langle \alpha_f J_f  || \hat O(E2) || \alpha_i J_i \rangle |^2 \label{BE2},
\end{equation}
where $J_i$, $J_f$ are the initial (final) angular momenta and $\alpha_i$ ($\alpha _f$) are other quantum numbers of the initial (final)
states.
The standard definition of a quadrupole moment in a state $| \alpha J \rangle $  reads
\begin{equation}
Q(J) =  \sqrt{ \dfrac{16 \pi }{5} }\langle \alpha J, M_J=J| \hat O(E2,M_L=0) | \alpha J, M_J=J \rangle \label{Q}.
\end{equation}

\begin{table}[t!] 
\tbl{Root-mean-square (rms) differences between the results on $E2$ and $M1$ transitions and 
moments from valence-space calculations  with either the state-dependent (st-dep) 
or the average parameterizations of the effective $E2$ or effective $M1$ operator and the respective NCSM results. 
See text for details.} 
{\begin{tabular}{@{}lcccccc@{}} \toprule
 & \multicolumn{2}{c}{$^{18}$O} & \multicolumn{2}{c}{$^{18}$F} & \multicolumn{2}{c}{$^{18}$Ne} \\  
 rms difference & st-dep & average & st-dep & average & st-dep & average \\ \colrule
 Red $E2$ M.E. [$ e\cdot\rm fm^2$] 	& 0.072 & 0.168 & 0.114 & 0.335 & 0.221 & 0.267 \\ \colrule 
 $B(E2)$ [$ e^2\cdot\rm fm^4$] 		& 0.017 & 0.039 & 0.189 & 0.240 & 0.208 & 0.583 \\ \colrule 
 $Q$     [$e\cdot\rm fm^2$]    	 	& 0.063 & 0.118 & 0.370 & 0.478 & 0.064 & 0.443  \\\botrule
Red $M1$ m.e. [$\mu_N$]    		& 0.070 & 0.088 & 0.050 & 0.071 & 0.108 & 0.118 \\ \colrule 
 $B(M1)$ [$\mu_N^2$]    		& 0.063 & 0.081 & 0.092 & 0.106 & 0.060 & 0.085 \\ \colrule 
 $\mu $  [$\mu_N$]      		& 0.023 & 0.032 & 0.189 & 0.182 & 0.020 & 0.099  \\\botrule 
\end{tabular} 
} 
\label{tab:rms} 
\end{table}

\begin{table}[t!] 
\tbl{$B(E2)$ values for transitions among the lowest states and electric quadrupole moments of the lowest states in $^{18}$O,
$^{18}$F and $^{18}$Ne: NCSM versus valence-space results with either state-dependent
or average effective charges. The experimental values are shown when available 
(only the latest result is retained in the table below
when a few experimental values are given in Ref.~\cite{Stone2005}).} 
{\begin{tabular}{lcccc} \toprule
                        & Exp~\cite{nndc,Stone2005} & NCSM & \multicolumn{2}{c}{$sd$ shell}  \\  
                        & 	 &      & st-dep & average \\ \colrule
 $^{18}$O               &     &      &    &          \\ 
 $B(E2; 2^+_1 \to 0^+_1)$ [$e^2\cdot\rm fm^4$] & 9.3(3) &   0.527 &   0.535 &  0.634  \\ \colrule 
 $B(E2; 2^+_2 \to 0^+_1)$ [$e^2\cdot\rm fm^4$] & 3.6(6) &   0.001 &   0.001 &  0.001  \\ \colrule 
 $B(E2; 4^+_1 \to 2^+_1)$ [$ e^2\cdot\rm fm^4$] & 3.3(2) &   0.339 &   0.359 &  0.434  \\ \colrule 
 $Q(2^+_1)$ [$e\cdot\rm fm^2$]              & $-3.6(9)$ & $-1.40$ & $-1.46$ & $-1.55$ \\ \botrule
 $^{18}$Ne               &     &      &        &          \\ 
 $B(E2; 2^+_1 \to 0^+_1)$ [$e^2\cdot\rm fm^4$] & 49.6(50) &   21.94 &   21.76 & 23.79  \\ \colrule 
 $B(E2; 2^+_2 \to 0^+_1)$ [$ e^2\cdot\rm fm^4$] & 1.9(9)   &   0.07  &    0.07 & 0.007  \\ \colrule 
 $B(E2; 4^+_1 \to 2^+_1)$ [$e^2\cdot\rm fm^4$] & 24.9(34) &   14.43 &   14.42 & 16.07  \\ \colrule 
 $Q(2^+_1)$ [$ e\cdot\rm fm^2$]                 &          & $-9.16$ & $-9.23$ & $-9.65$    \\\botrule
 $^{18}$F               &     &      &        &          \\ 
 $B(E2; 3^+_1 \to 1^+_1)$ [$e^2\cdot\rm fm^4$] & 16.3(6)      &   8.29  &  8.51 & 9.05  \\ \colrule 
 $B(E2; 3^+_2 \to 1^+_1)$ [$e^2\cdot\rm fm^4$] & 1.9(3)       &   0.014 & 0.010 & 0.030 \\ \colrule 
 $B(E2; 5^+_1 \to 3^+_1)$ [$e^2\cdot\rm fm^4$] & 16.2(7)      &   7.81  &  7.95 & 8.64  \\ \colrule 
 $Q(1^+_1)$ [$e\cdot\rm fm^2$]                 &              &   0.32  &  0.19 & 0.28  \\ \colrule
 $Q(3^+_1)$ [$e\cdot\rm fm^2$]                 &              & $-5.75$ & $-5.93$ & $-6.36$ \\\colrule
 $Q(5^+_1)$ [$ e\cdot\rm fm^2$]                & $\pm $7.7(5) & $-7.19$ & $-7.34$ & $-8.29$  \\\botrule
\end{tabular} 
} 
\label{tab:E2_comparison} 
\end{table}

We have calculated 66 $E2$ transitions in $^{18}$O, 66 $E2$ transitions in $^{18}$Ne and 269 $E2$ transitions in $^{18}$F within
the NCSM at $N_{\rm max}=4$ using the bare $E2$ operator and in the $sd$-shell model space using the effective operators.
The results are compared in Figs.~\ref{fig:BE2} and \ref{fig:Q} where we present absolute values of reduced $E2$ matrix elements and signed 
electric quadrupole moments for different states obtained in the $sd$-shell valence space calculations with the effective operators versus those obtained by
the NCSM. We observe an excellent correlation between the NCSM and valence-space values
with the root-mean-square (rms) differences between them 
being very small, especially for the state-dependent 
parameterization of the operator, as can be seen from Table~\ref{tab:rms}.
The $B(E2)$ values, used to calculate the rms difference, have been obtained for $J_i \ge J_f$.
Table~\ref{tab:E2_comparison} shows $B(E2)$ values for transitions between the lowest states and 
electric quadrupole moments for the lowest $J\ne 0$ states in $^{18}$O, $^{18}$Ne and  $^{18}$F
in comparison with the experimental data, when available.
We clearly see missing strength in the $E2$ transitions between the lowest $T=1$ states, which may signify a need for larger $N_{\rm max}$ values,
because of the very slow convergence of $E2$ observables~\cite{Shin2017}.
At the same time, the relative order of magnitude of theoretical $B(E2)$ values qualitatively resemble experimental findings.
In particular, our calculation suggests that the quadrupole moment of the $5^+_1$ state in $^{18}$F is negative.

It is worth noting here that the agreement is  worse  for a couple of cases. 
In particular, one of them is seen in Fig.~\ref{fig:Q} where
the quadrupole moment of the $1^+_7$ state
 in $^{18}$F with ${Q=0.78~e\cdot\rm fm^2}$ from the NCSM and ${Q=2.54~e\cdot\rm fm^2}$ from the valence-space calculations.
The discrepancy is most likely related to the fact that this state is the highest lying $1^+$ state (at around 25 MeV excitation energy) 
which has only 30\% of the valence-space component in the NCSM calculations. 
Without this point, the rms error for $Q$ in $^{18}$F reduces to from 0.370 to $0.114~e\cdot\rm fm^2$
for the state-dependent and from 0.478 to $0.339~e\cdot\rm fm^2$ for the  average parameterization of the operator.

\subsection{Effective $M1$ operator}

We proceed in a like manner for the effective $M1$ operator. 
From comparison of the NCSM results on magnetic dipole moments and transitions for $A=17$ at $N_{\rm max}=4$ with single-particle values, 
we have deduced state-dependent matrix elements of the orbital and spin contribution of the $M1$ operator
as well as respective state-dependent effective $g$-factors $g^{j}_{k}(a,b)$, $k=n,p$ and $j=l,s$. 
The rules can be summarized as follows:
\begin{equation}
\begin{array}{ll}
g^s_n(a,b) \langle b|| \mu_N \, \vec{s} \, || a \rangle  = \langle J_f|| \hat O(M1)_{\rm spin}  || J_i \rangle  & ({\rm from} \; ^{17}{\rm O}) \\ 
g^l_n(a,b) \langle b|| \mu_N \, \vec{l} \, || a \rangle  = \langle J_f|| \hat O(M1)_{\rm orbit} || J_i \rangle  & ({\rm from} \; ^{17}{\rm O}) \\ 
g^s_p(a,b) \langle b|| \mu_N \, \vec{s} \, || a \rangle  = \langle J_f|| \hat O(M1)_{\rm spin}  || J_i \rangle  & ({\rm from} \; ^{17}{\rm F}) \\ 
g^l_p(a,b) \langle b|| \mu_N \, \vec{l} \, || a \rangle  = \langle J_f|| \hat O(M1)_{\rm orbit} || J_i \rangle  & ({\rm from} \; ^{17}{\rm F}) \\ 
\end{array}
\end{equation}  
where spin and orbital part are referring to the first and second term of the $M1$ operator, given by Eq.~(\ref{M1}),
while $J_i$ and $J_f$ are the initial and final states obtained within the NCSM which are mapped onto the single-particle states $a$ and $b$ in the valence space.
Note that these equations help to exactly reproduce $M1$ matrix elements between the NCSM states for $A=17$ within the valence space, except for those
which involve the $s_{1/2}$ orbital due to the fact that it is characterized by $l=0$ orbital quantum number and therefore there is no way to reproduce
a small, but non-zero value of $\langle J_f|| \hat O(M1)_{\rm orbit} || J_i \rangle $. To overcome this difficulty, we can require that
in the case when $a$ or $b$ refers to the $s_{1/2}$ single-particle state, the valence space matrix element due to non-zero spin part would exactly
reproduce the NCSM matrix element of the full $M1$ operator. Because of the smallness of the orbital contribution to the NCSM results, 
both definitions lead to almost identical results for magnetic moments and, therefore, their use is practically equivalent.
We choose the latter method in the present work.
The corresponding effective $g$-factors are
summarized in Table~\ref{tab:eff_charges_gfactors}. We notice that there is a more modest quenching of the spin operators 
than the quenching  obtained in the previous studies in the $sd$-shell~\cite{Arima1987,Towner1987,Brown1987, USDab_em}. 
We notice also that the modifications to orbital $g$-factors found (positive for neutrons and negative for protons)
are opposite in sign from the corresponding corrections found in the same cited works.
For reference, the best fitted empirical values found for the same form of the $M1$ operator using the USDB wave functions
in the $sd$ shell are
$g^s_n=-3.55(10)$, $g^l_n=-0.09(26)$,  $g^s_p=5.15(9)$, $g^l_p=1.159(23)$~\cite{USDab_em}.

 
\begin{figure}[!] 
     \centerline{\includegraphics[width=3.7in]{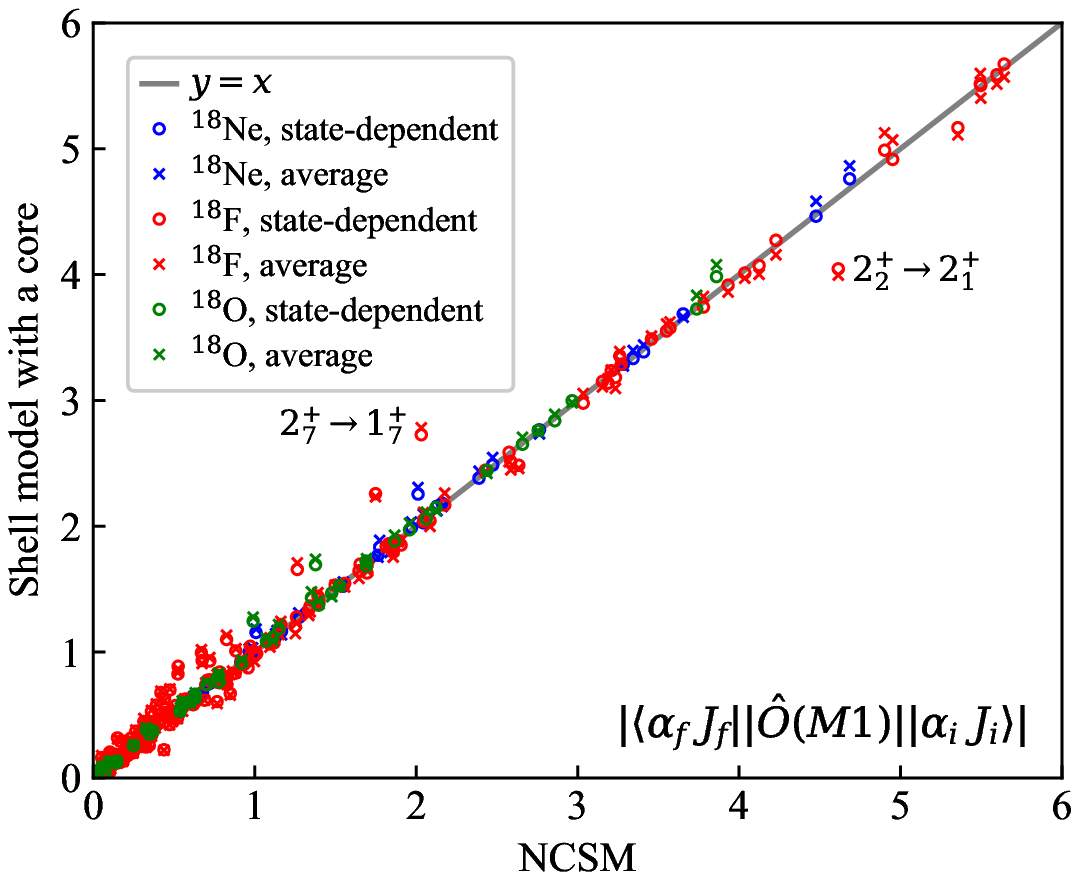}\label{fig:BM1_full}} 
    \centerline{\includegraphics[width=3.7in]{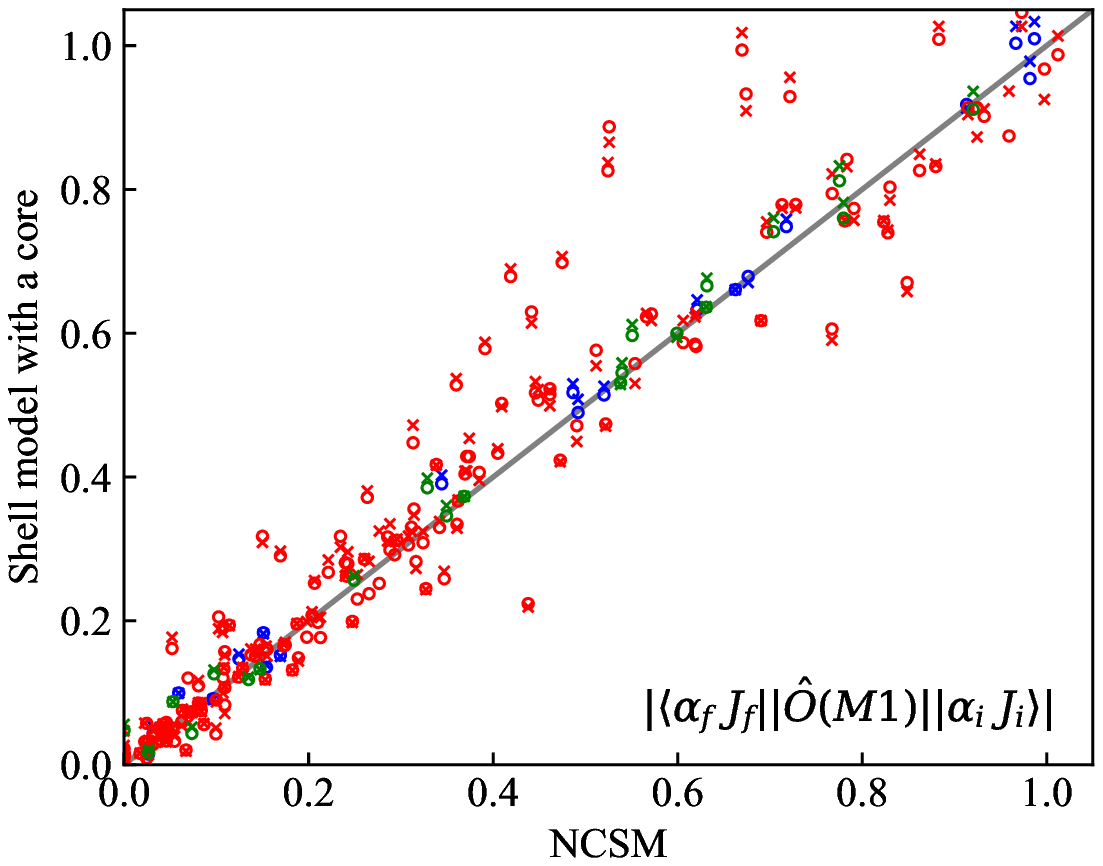}\label{fig:BM1_lim}} 
\caption{Comparison of the $sd$-shell results for the reduced $M1$ matrix elements between 
various states in $^{18}$O, $^{18}$F and $^{18}$Ne obtained using effective $M1$ operators 
within shell model with a core with 
those obtained by the NCSM (top); the same in the vicinity of zero  
in order to provide details on an enlarged scale for the weaker transitions (bottom).
Two outliers in $^{18}$F are flagged in the top panel with their state identifications and are discussed in the text.}\label{fig:BM1} 
\end{figure}

\begin{figure}[t] 
\centerline{ 
     {\includegraphics[width=3.7in]{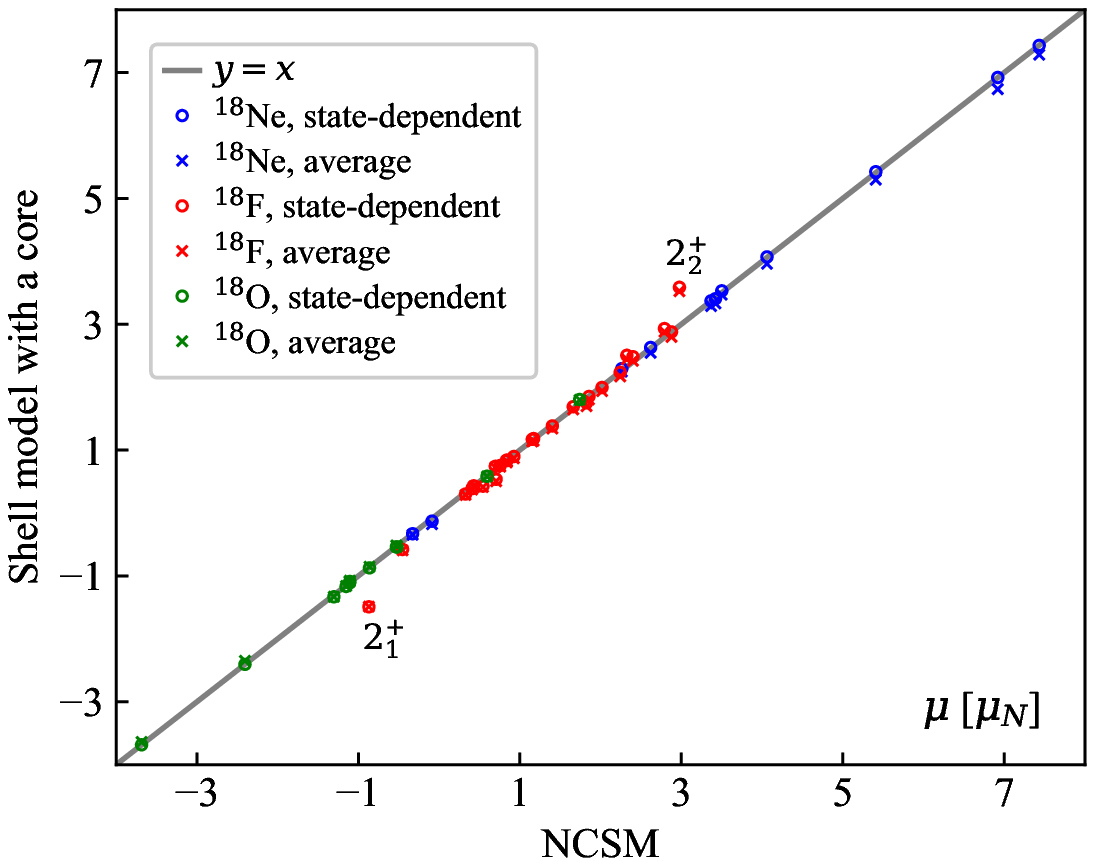}}
} 
\caption{Comparison of the $sd$-shell results for 
magnetic dipole moments of various states in $^{18}$O,
$^{18}$F and $^{18}$Ne obtained using effective $M1$ operators with those obtained by the NCSM.
Two outliers in $^{18}$F are flagged with their state identifications and are discussed in the text.}\label{fig:mu} 
\end{figure}


As a next step, we have calculated reduced $M1$ transition probabilities and magnetic moments in $A=18$ $sd$-shell nuclei. 
A reduced $M1$ transition matrix element, $ \langle \alpha_f J_f  || \hat O(M1) || \alpha_i J_i \rangle $, is required
to express the corresponding $B(M1)$ value of a transition from
an initial state $| \alpha_i J_i \rangle $ to the final state $| \alpha_f J_f \rangle $   as
\begin{equation}
B(M1;J_i \to J_f) =  \dfrac{1}{2J_i+1} |\langle \alpha_f J_f  || \hat O(M1) || \alpha_i J_i \rangle |^2 \label{BM1},
\end{equation}
while the magnetic moment reads
\begin{equation}
\mu(J) =  \sqrt{ \dfrac{4 \pi }{3} }\langle \alpha J, M_J=J| \hat O(M1,M_L=0) | \alpha J, M_J=J \rangle \label{mu}.
\end{equation}
The absolute values of reduced $M1$ matrix elements of 43 $M1$ transitions in $^{18}$O, 
43 $M1$ transitions in $^{18}$Ne and 212 $M1$ transitions in $^{18}$F obtained within
the NCSM at ${N_{\rm max}=4}$ using the bare $M1$ operator, and in the $sd$ shell model space using the effective operators
are compared in Fig.~\ref{fig:BM1}. The magnetic moments of all $J \ne 0$ states 
(11 states in  $^{18}$O,  11 states in $^{18}$Ne and 25 states in  $^{18}$F) are compared
in  Fig.~\ref{fig:mu}.
We notice again a remarkable agreement between the two sets of results. The state-dependent 
parameterization of the $M1$ single-particle matrix elements leads to  smaller rms differences
as compared to those provided by the average parameterization with an exception
of the $^{18}$F magnetic moments where both parameterizations provide nearly the same rms differences
(see Table~\ref{tab:rms}). The $B(M1)$ values, used to compute the rms difference, have been generated for $J_i \ge J_f$.

Table~\ref{tab:M1-comparison} shows $B(M1)$ values for transitions between the lowest states and magnetic dipole moments 
of the lowest states in $^{18}$O, $^{18}$Ne and  $^{18}$F
in comparison with the experimental data, when available.
In particular, there is a good agreement with the measured magnetic moment values.
We attribute this to the fact that the $M1$-observables converge faster in the NCSM calculations.

Two outlying $^{18}$F points are clearly seen in Fig.~\ref{fig:mu} and these are magnetic moments of the first
two $2^+$ states: they are only 130~keV separated in energy and appear to be mixed in isospin. 
The $2^+_1$ state contains 88\% of $T=1$ component and 11\% $T=0$ component and vice verse for $2^+_2$.
The values of the reduced $M1$ transition matrix elements between these two states are also significantly different between
the NCSM (4.258 $\mu_N^2$) and valence space calculations (3.297 $\mu_N^2$) as seen from Fig.~\ref{fig:BM1}.


\begin{table}[t!] 
\tbl{$B(M1)$ values for transitions between the lowest states and magnetic moments of the lowest states
in $^{18}$O, $^{18}$F and $^{18}$Ne: NCSM versus valence-space results obtained using either state-dependent or average $g$-factors.
The experimental values are shown when available (only the latest result is retained in the table below
when a few experimental values are given in Ref.~\cite{Stone2005}).}
{\begin{tabular}{lcccc} \toprule
                        & Exp~\cite{nndc,Stone2005}  & NCSM & \multicolumn{2}{c}{$sd$ shell}  \\  
                                      &              &         & st-dep  & average \\ \colrule
 $^{18}$O                             &              &         &         &          \\ 
 $B(M1; 2^+_2 \to 2^+_1)$ [$\mu_N^2$] & 0.25(4)      &   0.072 &   0.072 &  0.071  \\ \colrule 
 $B(M1; 1^+_1 \to 0^+_1)$ [$\mu_N^2$] &              &   0.101 &   0.119 &  0.125  \\ \colrule 
 $\mu(2^+_1)$ [$\mu_N$]               & $-0.57(3)$   & $-0.53$ & $-0.54$ & $-0.52$ \\ \colrule
 $\mu(4^+_1)$ [$\mu_N$]               & $\pm 2.5(4)$ & $-2.41$ & $-2.41$ & $-2.35$ \\ \botrule
 $^{18}$Ne                            &              &         &         &          \\ 
 $B(M1; 2^+_2 \to 2^+_1)$ [$\mu_N^2$] &              &   0.092 &   0.092 &  0.090  \\ \colrule 
 $B(M1; 1^+_1 \to 0^+_1)$ [$\mu_N^2$] &              &   0.079 &   0.089 &  0.093  \\ \colrule 
 $\mu(2^+_1)$ [$\mu_N$]               &              &   2.62  &   2.63  & 2.55 \\ \botrule 
 $^{18}$F                             &              &         &         &          \\ 
 $B(M1; 1^+_1 \to 0^+_1)$ [$\mu_N^2$] &              &   5.428 &  5.366  &  5.256  \\ \colrule 
 $\mu(3^+_1)$ [$\mu_N$]               & $+1.77(12)$  &   1.857 &  1.852  &  1.803 \\ \botrule
 $\mu(5^+_1)$ [$\mu_N$]               & $+2.86(3)$   &   2.878 &  2.878  &  2.798 \\ \botrule
\end{tabular} 
} 
\label{tab:M1-comparison} 
\end{table}

\section{Conclusions and Prospects} 

In this contribution we present a novel method for deriving effective one-body electromagnetic operators for
valence space calculations from the NCSM results. 
The effective $E2$ and $M1$ one-body operators are obtained as a set of parameterized effective single-particle matrix elements
to be used with many-body states in the valence space.
Comparison of a large number of reduced $E2$ and $M1$ matrix elements, as well as electric quadrupole and 
magnetic dipole moments
in $A=18$ $sd$-shell nuclei shows a very good agreement with the NCSM values.
Further exploration of these operators in systems with more valence particles is in progress.
One also anticipates systematically increasing the space for the NCSM calculations as computational resources allow.

\section*{Acknowledgments}

Z. Li and N.~A.~Smirnova acknowledge the financial support of CNRS/IN2P3 via ENFIA Master project, France. 
The work of A.~M.~Shirokov is supported by the Russian Foundation for Basic Research under Grant No. 20-02-00357.
The work of I.~J.~Shin was
supported by the Rare Isotope Science Project of Institute for Basic Science funded by Ministry of Science
and ICT and National Research Foundation of Korea (2013M7A1A1075764).
This work was supported in part by the US Department of Energy (DOE) under Grant Nos. DE-FG02-87ER40371,
DE-SC0018223 (SciDAC-4/NUCLEI). 
Computational resources were provided by the National Energy Research Scientific Computing Center (NERSC), 
which is supported by the US DOE Office of Science under Contract No. DE-AC02-05CH11231,
and also provided by the National Supercomputing Center, Republic of Korea, including technical support (KSC-2020-CRE-0027).
N.~A.~Smirnova and A.M. Shirokov thank the Institute for Basic Science, Daejeon, for a hospitality and financial support of their visits.



 
 
\bibliographystyle{ws-rv-van} 

\bibliography{Eff-op}

 
\end{document}